\overfullrule=0pt
\input harvmac
\input amssym.tex
\def\a{{\alpha}}

\def\ab{{\overline\alpha}}

\def\l{{\lambda}}
\def\lb{{\overline\lambda}}

\def\b{{\beta}}
\def\bb{{\overline\beta}}

\def\g{{\gamma}}
\def\gb{{\overline\gamma}}

\def\d{{\delta}}
\def\db{{\overline\delta}}

\def\r{{\rho}}
\def\rb{{\overline\rho}}

\def\N{{\nabla}}
\def\Nb{{\overline\nabla}}
\def\Nh{{\widehat\nabla}}
\def\O{{\Omega}}

\def\Ob{{\overline\O}}
\def\Oh{{\widehat\O}}
\def\o{{\omega}}
\def\ob{{\overline\omega}}

\def\p{{\partial}}
\def\pb{{\overline\partial}}
\def\t{{\theta}}

\def\L{{\Lambda}}
\def\Lb{{\overline\Lambda}}

\def\Pib{{\overline\Pi}}
\def\Jb{{\overline J}}

\def\S{{\Sigma}}

\def\Th{{\widehat T}}

\def\AB{{\bf A}}
\def\WB{{\bf W}}
\def\FB{{\bf F}}
\def\WW{{\overline W}}
\def\pp{{\overline p}}
\def\tb{{\overline\theta}}
\def\dd{{\overline d}}

\def\CC{{\overline C}}
\def\Psib{{\overline\Psi}}
\def\taub{{\overline\tau}}
\def\NN{{\overline N}}
\def\PB{{\bf P}}
\def\SB{{\bf S}}

\def\EB{{\bf E}}
\def\EBB{{\bf\overline\EB}}
\def\OB{{\bf\O}}
\def\OBB{{\bf\overline\O}}
\def\CB{{\bf C}}
\def\CBB{{\bf\overline C}}
\def\SS{{\overline S}}
\def\Phib{{\bf\Phi}}
\def\MB{{\bf M}}
\def\LB{{\bf\L}}
\def\LBB{{\bf\overline\L}}

\baselineskip 12pt

\Title{ \vbox{\baselineskip 12pt
}}
{\vbox{\centerline
{ General fluctuations of the type II pure spinor   }
\bigskip
\centerline{ string on curved backgrounds }
}}
\smallskip
\centerline{Osvaldo Chandia\foot{e-mail: ochandiaq@gmail.com}, }
\smallskip
\centerline{\it Departamento de Ciencias, Facultad de Artes Liberales, Universidad Adolfo Ib\'a\~nez}
\centerline{\it \& UAI Physics Center, Universidad Adolfo Ib\'a\~nez}
\centerline{\it Diagonal Las Torres 2640, Pe\~nalol\'en, Santiago, Chile} 

\bigskip
\bigskip

\noindent
The general fluctuations, in the form of vertex operators, for the type II superstring in the pure spinor formalism are considered. We review the construction of these vertex operators in flat space-time. We then review the type II superstrings in curved background in the pure spinor formalism to finally construct the vertex operators on a generic type II supergravity background.

\Date{March 2019}


\lref\BerkovitsFE{
  N.~Berkovits, 
  ``Super Poincare covariant quantization of the superstring,''
JHEP {\bf 0004}, 018 (2000).
[hep-th/0001035].
}

\lref\BerkovitsUE{
  N.~Berkovits and P.~S.~Howe,
  ``Ten-dimensional supergravity constraints from the pure spinor formalism for the superstring,''
Nucl.\ Phys.\ B {\bf 635}, 75 (2002).
[hep-th/0112160].
}

\lref\ChandiaHN{
  O.~Chandia and B.~C.~Vallilo,
  ``Conformal invariance of the pure spinor superstring in a curved background,''
JHEP {\bf 0404}, 041 (2004).
[hep-th/0401226].
}

\lref\BedoyaIC{
  O.~A.~Bedoya and O.~Chandia,
  ``One-loop Conformal Invariance of the Type II Pure Spinor Superstring in a Curved Background,''
JHEP {\bf 0701}, 042 (2007).
[hep-th/0609161].
}

\lref\ChandiaVP{
  O.~Chandia and M.~Tonin,
  ``BRST anomaly and superspace constraints of the pure spinor heterotic string in a curved background,''
JHEP {\bf 0709}, 016 (2007).
[arXiv:0707.0654 [hep-th]].
}

\lref\BerkovitsYR{
  N.~Berkovits and O.~Chandia,
  ``Superstring vertex operators in an $AdS_5\times S^5$ background,''
Nucl.\ Phys.\ B {\bf 596}, 185 (2001).
[hep-th/0009168].
}

\lref\ChandiaEKC{
  O.~Chandia and B.~C.~Vallilo,
  ``Vertex operators for the plane wave pure spinor string,''
JHEP {\bf 1810}, 088 (2018).
[arXiv:1807.05149 [hep-th]].
}

\lref\ChandiaNYH{
  O.~Chandia,
  ``General fluctuations of the heterotic pure spinor string on curved backgrounds,''
[arXiv:1812.05124 [hep-th]].
}

\lref\ChandiaAFC{
  O.~Chandia and B.~C.~Vallilo,
  ``A superfield realization of the integrated vertex operator in an $AdS_5\times S^5$ background,''
JHEP {\bf 1710}, 178 (2017).
[arXiv:1709.05517 [hep-th]].
}

\lref\BerkovitsBT{
  N.~Berkovits,
  ``Pure spinor formalism as an N=2 topological string,''
JHEP {\bf 0510}, 089 (2005).
[hep-th/0509120].
}

\lref\ChandiaIX{
  O.~Chandia,
  ``The $b$ Ghost of the Pure Spinor Formalism is Nilpotent,''
Phys.\ Lett.\ B {\bf 695}, 312 (2011).
[arXiv:1008.1778 [hep-th]].
}

\lref\JusinskasYCA{
  R.~Lipinski Jusinskas,
  ``Nilpotency of the b ghost in the non-minimal pure spinor formalism,''
JHEP {\bf 1305}, 048 (2013).
[arXiv:1303.3966 [hep-th]].
}

\lref\ChandiaSTA{
  O.~Chandia and B.~C.~Vallilo,
  ``Non-minimal fields of the pure spinor string in general curved backgrounds,''
JHEP {\bf 1502}, 092 (2015).
[arXiv:1412.1030 [hep-th]].
}

\lref\ChandiaIMA{
  O.~Chandia,
  ``The Non-minimal Heterotic Pure Spinor String in a Curved Background,''
JHEP {\bf 1403}, 095 (2014).
[arXiv:1311.7012 [hep-th]].
}

\lref\BerkovitsAMA{
  N.~Berkovits and O.~Chandia,
  ``Simplified Pure Spinor $b$ Ghost in a Curved Heterotic Superstring Background,''
JHEP {\bf 1406}, 001 (2014).
[arXiv:1403.2429 [hep-th]].
}

\lref\ChandiaKJA{
  O.~Chandia, A.~Mikhailov and B.~C.~Vallilo,
 ``A construction of integrated vertex operator in the pure spinor sigma-model in $AdS_5 \times S^5$,''
JHEP {\bf 1311}, 124 (2013).
[arXiv:1306.0145 [hep-th]].
}

\newsec{Introduction}

The pure spinor formalism was invented nineteen years ago with the idea of provide a quantizable string sigma-model on any background space-time geometry \BerkovitsFE. The pure spinor formalism is manifestly space-time supersymmetric and contains a world-sheet field, known as pure spinor, which has two main utilities. It allows to have a conformal invariant system and allows quantization through the existence of an operator that has the properties of a BRST operator. Soon after, the explicit form of the superstrings on curved background was studied in \BerkovitsUE. Here the classical BRST invariance of the world-sheet action puts the background to satisfy the equations of ten-dimensional supergravity. It was also proven that the world-sheet 
system is conformal invariant at the one-loop quantum level, for the heterotic superstring \ChandiaHN\ and for the type II superstring \BedoyaIC. The quantum local symmetries were also studied in \ChandiaVP. 

The fluctuations around the background are BRST invariant and describe the physical content of the string theory. For example, the open string massless fluctuations are described by the unintegrated vertex operator $U=\l^\a \AB_\a(X,\t)$ where the superfield $\AB$ depends on the ten-dimensional superspace coordinates $X$ and $\t$. BRST invariance of $U$ gives the equations of super-Maxwell for $\AB_\a$, that is, it describes the photon and the photino in ten dimensions. Note that only one operator is enough to describe the physical massless states of the string theory, unlike in RNS where there are different operators for photon and the photino. A similar analysis can be done for other superstrings, we will review the type II case below. All this is done in a flat space-time background. We could ask for the equivalent of $U$ in non-flat backgrounds. This was done in \BerkovitsYR\ for the $AdS_5\times S^5$ background, its plane-wave limit in \ChandiaEKC\  and in for the heterotic string in \ChandiaNYH\ a generic background. Our goal is to study the case of type II in a generic supergravity background. 

Besides unintegrated vertex operators, it is possible to study integrated vertex operators. They are fluctuations of the world-sheet action and can be used to compute scattering amplitudes for $4$ or more states. The integrated vertex operator is obtained from the unintegrated vertex operator through a descent procedure. For the open string in flat background space-time, $\p U$ is BRST trivial, then there exists an operator $V$ that satisfies $QV=\p U$. This is the integrated vertex operator. This procedure has to work, because there exists a $b$ ghost in the theory, although this $b$ ghost is a composite operator in the pure spinor formalism. To have the descent working correctly, it is necessary that $b$ is conserved, nilpotent and satisfies $Qb=T$ where $T$ is the stress-energy tensor.

The integrated vertex operator for type II superstring in a flat space-time background can be constructed in this manner and we review this in section $2$. A similar construction was done in \ChandiaAFC\ for the $AdS_5\times S^5$ background, its plane-wave limit in \ChandiaEKC\ and for the heterotic string in a generic background in \ChandiaNYH. Our purpose is to generalize this result for the type II superstring in a generic supergravity background.

This paper is organized as follows. We review the type II superstring in a flat space-time background in section $2$. In the section $3$, we review the type II superstring in a curved background. In particular the necessary BRST transformations of the world-sheet fields are discussed. In section $4$ we study the unintegrated vertex operator in a curved background, we obtain a chain of superfields by applying successive superspace covariant derivatives on the superfield of the unintegrated vertex operator. In section $5$ we perform the descent procedure to determine the integrated vertex operator in a curved supergravity background. We end in section $6$ with some final comments.

\newsec{Review of the type II superstring in a flat background}

We now review the type II superstring in a flat ten-dimensional background using the pure spinor formalism. The action is given by
\eqn\sflat{ S= \int d^2z ~ \ha\p X^m \pb X_m + p_\a \pb \t^\a +\pp_\ab \p \tb^\ab  + \o_\a \pb \l^\a + \ob_\ab \p \lb^\ab ,}
where $(X^m,\t^\a,\tb^\ab)$ are the superspace coordinates in flat ten-dimensional space-time ($m=0, \dots, 9, \a,\ab=1, \dots, 16$), $(p_\a,\pp_\ab)$ are the momentum conjugate variables of $(\t^\a,\tb^\ab)$, $(\l^\a,\o_\a)$ and $(\lb^\ab,\ob_\ab)$ are the pure spinor conjugate pairs of variables. The pure spinor variables are constrained by the the so called pure spinor conditions $\l\g^m\l=\lb\g^m\lb=0$. The pure spinor conjugate variables are defined up to the gauge invariances $\d\o_\a=(\l\g^m)_\a\L_m, \d\ob_\ab=(\lb\g^m)_\a\Lb_m$, then only $44$ out of $64$ $(\l,\lb)$ and $(\o,\ob)$ variables are independent \BerkovitsFE. The left-moving central charge has the following contributions: $+10$ from $X$, $-32$ from $(p,\t)$ and $+22$ from $(\l,\o)$, so the total central charge is zero. Similarly, for the the right-moving central charge we have $+10$ from $X$, $-32$ from $(\pp,\tb)$ and $+22$ from $(\lb,\ob)$ adding up, again, to zero.  Therefore, the action \sflat\ is conformal invariant. However, this is not enough to describe the physical states of type II superstring. A nilpotent operator, which we call BRST operator, was proposed in \BerkovitsFE\ to describe the correct spectrum of the string theory. For type II, the BRST operator is
\eqn\Qz{Q=\oint \l^\a d_\a + \lb^\ab d_\ab ,}
where $d_\a$ and $\dd_\ab$ are the generators of superspace translations and are given by
\eqn\dsare{\eqalign{&d_\a=p_\a+\ha(\g^m\t)_\a\p X_m -{1\over8}(\g^m\t)_\a(\t\g^m\p\t),\cr &\dd_\ab=\pp_\ab+\ha(\g^m\tb)_\ab\pb X_m -{1\over8}(\g^m\tb)_\ab(\tb\g^m\pb\tb) .\cr}}
Note that $Q^2=0$ because the OPE's
\eqn\dsope{\eqalign{&d_\a(y) d_\b(z)\to-{1\over(y-z)}\g^m_{\a\b}\Pi_m(z) ,\cr &\dd_\ab({\bar y})\dd_\bb({\bar z}) \to -{1\over({\bar y}-{\bar z})} \g^m_{\ab\bb}\Pib_m({\bar z}) ,\cr}}
and the pure spinor conditions. Here $\Pi_m=\p X_m+\ha(\t\g_m\p\t)+\ha(\tb\g_m\p\tb)$ and~$\Pib_m=\pb X_m+\ha(\t\g_m\pb\t)+\ha(\tb\g_m\pb\tb)$. 

Physical states of the type II strings are in the cohomology of the BRST operator \Qz. The massless unintegrated vertex operator has ghost number $2$ (that is $1~\l$ and $1~\lb$) and it is given by
\eqn\Uflat{U=\l^\a\lb^\ab\AB_{\a\ab}(X,\t,\tb) .}
$QU=0$ implies the equations 
\eqn\QUflat{\l^\a\l^\b D_\a \AB_{\b\ab}=0,\quad \lb^\ab\lb^\bb D_{\ab}\AB_{\a\bb}=0 ,}
where 
$$D_\a={\p\over{\p\t^\a}}+\ha(\g^m\t)_\a{\p\over{\p X^m}},\quad D_\ab={\p\over{\p\tb^\ab}}+\ha(\g^m\tb)_\ab{\p\over{\p X^m}} .$$
The equations \QUflat\ imply the existence of the superfields $\AB_{m\ab}$ and $\AB_{\a m}$ defined by
\eqn\Ams{D_{(\a}\AB_{\b)\ab}=\g^m_{\a\b}\AB_{m\ab},\quad D_{(\ab}\AB_{\a\bb)}=\g^m_{\ab\bb}\AB_{\a m} .}
It is possible to construct covariant superfields by applying $D_\a$ and $D_\ab$ to these equations. As it was shown in \ChandiaAFC, we obtain the following defining relations
\eqn\DL{D_\a\AB_{m\ab}-\p_m\AB_{\a\ab}=(\g_m)_{\a\b}\WB^\b{}_\ab,\quad D_\a\WB^\b{}_{\ab}={1\over4}(\g^{mn})_\a{}^\b\FB_{mn\ab},}
\eqn\DR{D_\ab\AB_{\a m}-\p_m\AB_{\a\ab}=(\g_m)_{\ab\bb}\WB_\a{}^\bb,\quad D_\ab\WB_\a{}^{\bb}={1\over4}(\g^{mn})_\ab{}^\bb\FB_{\a mn},}
from \Ams. Here $\FB_{mn\ab}=\p_{[m}\AB_{n]\ab}, \FB_{\a mn}=\p_{[m}\AB_{\a n]}$.

These relations help to find the integrated vertex operator. It is given through the following descent procedure. $\p U$ and $\pb U$ are trivial because it is possible to find operators that satisfy $\p U=Q W$ and  $\pb U = Q\WW$. Actually, any field with non-zero conformal weight is BRST trivial. This statement is based on the existence of the conformal ghost $b$. Although the $b$ ghost is not a basic field in \sflat, it is possible to construct a composite field that satisfies $Qb=T$ \BerkovitsBT, the stress-energy tensor, and it is nilpotent \ChandiaIX\ \JusinskasYCA. Similarly, $\p\WW-\pb W$ is also trivial. In fact, there exists an operator $V$ that satisfies $QV=\p\WW-\pb W$. This is the integrated vertex operator. From the unintegrated vertex $U$ \Uflat, $W$ and $\WW$ are given by
\eqn\WWW{\eqalign{&W=\lb^\ab\left(\p\t^\a\AB_{\a\ab}+\Pi^m\AB_{m\ab}+d_\a\WB^\a{}_\ab+\ha N^{mn}\FB_{mn\ab} \right) ,\cr &\WW=\l^\a\left(\pb\tb^\ab\AB_{\a\ab}+\Pib^m\AB_{\a m}+\dd_\ab\WB_\a{}^\ab+\ha\NN^{mn}\FB_{\a mn}\right) ,\cr }}
where $N^{mn}=\ha(\l\g^{mn}\o)$ and $\NN^{mn}=\ha(\lb\g^{mn}\ob)$. The integrated vertex operator becomes
\eqn\ivoz{\eqalign{V&=\p\t^\a\pb\tb^\ab\AB_{\a\ab}+\p\t^\a\Pib^m\AB_{\a m}-\Pi^m\pb\tb^\ab\AB_{m\ab}+d_\a\pb\tb^\ab\WB^\a{}_\ab+\p\t^\a\dd_\ab\WB_\a{}^\ab+\ha\p\t^\a\NN^{mn}\FB_{\a mn} \cr &-\ha N^{mn}\pb\tb^\ab \FB_{mn\ab}+\Pi^m\Pib^n\AB_{mn} +\Pi^m\dd_\ab\EB_m{}^\ab+d_\a\Pib^m\EBB_m{}^\a+d_\a\dd_\ab \PB^{\a\ab}+\ha N^{mn}\Pib^p \OBB_{pmn}\cr& +\ha\Pi^p\NN^{mn}\OB_{pmn}+\ha d_\a \NN^{mn} \CBB_{mn}{}^\a+\ha N^{mn}\dd_\ab\CB_{mn}{}^\ab+{1\over4} N^{mn}\NN^{pq}\SB_{mnpq}   ,\cr }}
where the superfields $\AB_{mn}, \dots, \SB_{mnpq}$ are defined after taking higher $D_\a$ and $D_\ab$ superspace covariant derivatives. Namely,  
\eqn\more{\eqalign{D_{(\a} \AB_{\b)n}&=\g^m_{\a\b}\AB_{mn},\quad D_{(\ab} \AB_{m\bb)}=-\g^n_{\ab\bb}\AB_{mn},\cr D_{(\a} \WB_{\b)}{}^\ab&=-\g^m_{\a\b}\EB_m{}^\ab,\quad D_{(\ab}\WB^\a{}_{\bb)}=\g^m_{\ab\bb}\EBB_m{}^\a,\cr D_{(\a}\FB_{\b)mn}&=\g^p_{\a\b}\OB_{pmn},\quad D_{(\ab}\FB_{mn\bb)}=-\g^p_{\ab\bb}\OBB_{pmn},\cr D_\a\EB_m{}^\ab+\p_m\WB_\a{}^\ab&=-(\g_m)_{\a\b}\PB^{\b\ab},\quad D_\ab\EBB_m{}^\a-\p_m\WB^\a{}_\ab=(\g_m)_{\ab\bb}\PB^{\a\bb},\cr D_\a\OB_{pmn}-\p_p\FB_{\a mn}&=(\g_p)_{\a\b}\CBB_{mn}{}^\a ,\quad D_\ab\OBB_{pmn}+\p_p\FB_{mn\ab}=-(\g_p)_{\ab\bb}\CB_{mn}{}^\ab,\cr D_\a\CBB_{pq}{}^\a&={1\over4}(\g^{mn})_\a{}^\b\SB_{mnpq},\quad D_\ab\CB_{mn}{}^\bb={1\over4}(\g^{pq})_\ab{}^\bb\SB_{mnpq} .\cr}}
The physical interpretation of the these superfields is simple. In the $(\t,\tb)$ expansion, $\AB_{mn}=h_{mn}+\cdots$, where $h_{mn}$ describes the graviton, the Kalb-Ramond field and the dilaton. The superfields $\EB_m{}^\ab=\psi_m{}^\ab+\cdots, \EBB_m{}^\a={\overline\psi}_m{}^\a+\cdots$, where $\psi$ and 
${\overline\psi}$ are the gravitini and the dilatini. The superfield $\PB^{\a\ab}=f^{\a\bb}+\cdots$, where $f$ is the Ramond-Ramond field-strength. The other superfields in \ivoz\ have expansions with components related to $h, \psi, {\overline\psi}$ and $f$.

In the next section we will describe the regime beyond the linearized level where $\OB$ and $\OBB$ become the background Lorentz connections, $\EB$ and $\EBB$  are the vielbein superfields and $\SB$ becomes related to the superspace curvature. 

\newsec{Review of the type II superstring in a curved background}

The action for the type II superstring can be obtained by adding the integrated vertex operator \ivoz\ to flat string action \sflat\ and then covariantize respect to the bakcground supergeometry invariance \BerkovitsUE. Then, the sigma model action becomes
\eqn\SC{S= \int d^2z ~ \ha\Pi_a\Pib^a + \ha\Pi^A\Pib^B B_{BA}+d_\a\Pib^\a+\dd_\ab\Pi^\ab+\o_\a\Nb\l^\a+\ob_\ab\N\lb^\ab}
$$+d_\a\dd_\ab P^{\a\bb}+\l^\a\o_\b\dd_\gb C_\a{}^{\b\gb}+\lb^\ab\ob_\bb d_\g \CC_{\ab}{}^{\bb\g}+\l^\a\o_\b\lb^\gb\ob_\db S_{\a\gb}{}^{\b\db},$$
where $\Pi^A=\p Z^M E_M{}^A$ and $\Pib^A=\pb Z^M E_M{}^A$ with $E_M{}^A$ being the vielbein and $Z^M$ the curved superspace coordinates. Note that the fields coupled to $\o$ and $\ob$ has to respect the pure spinor gauge symmetry $\d\o_\a=(\l\g^a)_\a\L_a$ and $\d\ob_\ab=(\lb\g^a)_\ab\Lb_a$. Then, the forms of $C, \CC, S$ are
\eqn\CCS{\eqalign{&C_\a{}^{\b\gb}=\d_\a^\b C^\gb+{1\over4}(\g^{ab})_\a{}^\b C_{ab}{}^\gb,\quad \CC_\ab{}^{\bb\g}=\d_\ab^\bb\CC^\g+{1\over4}(\g^{ab})_\ab{}^\bb\CC_{ab}{}^\g,\cr&S_{\a\gb}{}^{\b\gb}=\d_\a^\b \d_\gb^\db S + {1\over4}(\g^{ab})_\a{}^\b \d_\gb^\bb S_{ab}+{1\over4}\d_\a^\b(\g^{ab})_\gb{}^\db\SS_{ab} + {1\over{16}}(\g^{ab})_\a{}^\b(\g^{cd})_\gb{}^\db S_{abcd}  .\cr }}
The covariant derivatives on the pure spinors variables are 
\eqn\covD{\Nb\l^\a=\pb\l^\a+\l^\b\Ob_\b{}^\a,\quad \N\lb^\ab=\p\lb^\ab+\lb^\bb\O_\bb{}^\ab ,}
where $\O_\ab{}^\bb=\p Z^M\O_{M\ab}{}^\bb$ and $\Ob_\a{}^\b=\pb Z^M\O_{M\a}{}^\b$ with $\O_M$ being the Lorentz connections. They are of the form \BerkovitsUE\ 
\eqn\conn{\O_{M\a}{}^\b=\d_\a^\b\O_M+{1\over4}(\g^{ab})_\a{}^\b\O_{Mab},\quad \O_{M\ab}{}^\bb=\d_\ab^\bb\Oh_M+{1\over4}(\g^{ab})_\ab{}^\bb\Oh_{Mab}, }
then there are two possible covariant derivatives acting on a vector. For example, we could have
\eqn\cova{\N\Pi^a=\p\Pi^a+\Pi^b\p Z^M\O_{Mb}{}^a,} 
or
\eqn\covb{\Nh\Pi^a=\p\Pi^a+\Pi^b\p Z^M\Oh_{Mb}{}^a.}
Fortunately, both covariant derivatives are related after solving the constraints on the background fields in \SC\ dictated by the BRST symmetry generated by 
$Q=\oint \left(\l^\a d_\a + \lb^\ab d_\ab\right)$. These constraints are expressed in terms of the components of the torsion $2$-form $T^A=dE^A+E^B\wedge\O_B{}^A$  (note that $T^a$ is defined with the connection $\O$ and $\Th^a$ is defined with the connection $\Oh$), the curvature $2$-form $R_A{}^B=d\O_A{}^B+\O_A{}^C\wedge\O_C{}^B$ and the $3$-form field-strength $H=dB$. The indices $A,B,C,\dots$ take values in $(a,\a,\ab)$. 
Note the $T_{AB}{}^a=\Th_{AB}{}^a$ for $(A,B)\in(\a,\ab)$. 

The constraints for some of the torsion, $H$ and curvature components derived in \BerkovitsUE\ are solved by 
\eqn\sola{T_{\a\b}{}^a=-\g^a_{\a\b},\quad T_{\ab\bb}{}^a=-\g^a_{\ab\bb},\quad R_{\a\b\gb}{}^\db=0,\quad R_{\ab\bb\g}{}^\d=0  ,}
\eqn\solb{T_{\a\b}{}^\g=T_{\a\bb}{}^\g=T_{\ab\bb}{}^\g=T_{\a\b}{}^\gb=T_{\a\bb}{}^\gb=T_{\ab\bb}{}^\gb=T_{a\a}{}^\b=T_{a\ab}{}^\bb=0 ,}
\eqn\solc{H_{a\a\b}=-(\g_a)_{\a\b},\quad H_{a\ab\bb}=(\g_a)_{\ab\bb},\quad H_{\a\bb A}=H_{\a\b\g}=H_{\ab\bb\gb}=H_{ab\a}=H_{ab\ab}=0 .}

The Bianchi identities help to find relations for the other $T, H$ and $R$ components. These identities are defined as
\eqn\bianchi{\eqalign{&(\N T)_{ABC}{}^D=\N_{[A}T_{BC]}{}^D+T_{[AB}{}^ET_{EC]}{}^D,\cr &(\N H)_{ABCD}=\N_{[A}H_{BCD]}+{3\over2}T_{[AB}{}^EH_{ECD]},\cr &(\N R)_{ABCD}{}^E=\N_{[A}R_{BC]D}{}^E+T_{[AB}{}^FR_{FC]D}{}^E .} }
Note again that there are two types of Bianchi identities for $D=d$ in the first relation of \bianchi\ depending if one uses $T^d$ or $\Th^d$. As it was shown in \BerkovitsUE\ (see also\BedoyaIC), these identities imply 
\eqn\torsiona{T_{\a a}{}^b=2(\g_a{}^b)_\a{}^\b\O_\b,\quad T_{\ab a}{}^b=0,\quad \O_\ab=\O_a=0 ,}
\eqn\torsionb{\Th_{\ab a}{}^b=2(\g_a{}^b)_\ab{}^\bb\Oh_\bb,\quad \Th_{\a a}{}^b=0,\quad \Oh_\a=\Oh_a=0 ,}
\eqn\torsionc{H_{abc}=-T_{abc}=\Th_{abc} .}
Using these results, the $\Oh_{ab}$ one-form connection is known in terms of the other connections. In fact, we obtain
\eqn\OMhat{\Oh_{cab}=\O_{cab}-T_{cab},\quad \Oh_{\a ab}=\O_{\a ab}-T_{\a ab},\quad \Oh_{\ab ab}=\O_{\ab ab}+\Th_{\ab ab} .}
From now on we will use the connection $\O_{ab}$ in the covariant derivatives.

The background fields $P, C, \CC, S$ satisfy the relations
\eqn\hol{T_{a\a}{}^\bb=(\g_a)_{\a\g}P^{\g\bb},\quad R_{a\a\bb}{}^\gb=-(\g_a)_{\a\d}\CC_\bb{}^{\gb\d},\quad C_\a{}^{\b\gb}=-\N_\a P^{\b\gb},}
$$S_{\a\gb}{}^{\b\db}=\N_\a\CC_\gb{}^{\db\b}+R_{\a\rb\gb}{}^\db P^{\b\rb} ,$$
and 
\eqn\ahol{T_{a\ab}{}^\b=-(\g_a)_{\ab\gb}P^{\b\gb},\quad R_{a\ab\b}{}^\g=-(\g_a)_{\ab\db}C_\b{}^{\g\db},\quad \CC_\ab{}^{\bb\g}=\N_\ab P^{\g\bb} ,}
$$S_{\a\gb}{}^{\b\db}=\N_\gb C_\a{}^{\b\db}-R_{\gb\r\a}{}^\b P^{\r\db} .$$
Note the expressions for $S$ are equivalent because if one subtract both equations and uses the first lines of \hol\ and \ahol\ one obtains zero.

The BRST transformations of the world-sheet fields were obtained in \ChandiaSTA\ and, up to a Lorentz transformation with field-dependent parameters, are given by 
\eqn\QC{\eqalign{&Q\l^\a=Q\lb^\ab=0,\quad Q\o_\a=d_\a,\quad Q\ob_\ab=\dd_\ab, \cr
&Qd_\a=-(\g_a\l)_\a\Pi^a +\left(\l^\b R_{\a\b\g}{}^\d+\lb^\bb R_{\a\bb\g}{}^\d\right)\l^\g\o_\d  ,\cr
&Q\dd_\ab=(\g_a\lb)_\ab\Pib^a+\left(\l^\b R_{\ab\b\gb}{}^\db+\lb^\bb R_{\ab\bb\gb}{}^\db\right)\lb^\gb\ob_\db ,\cr
&Q\Pi^a=-\l^\a\Pi^AT_{A\a}{}^a-\lb^\ab\Pi^AT_{A\ab}{}^a,\quad   Q\Pib^a=-\l^\a\Pib^AT_{A\a}{}^a-\lb^\ab\Pib^AT_{A\ab}{}^a,\cr
& Q\Pi^\a=\N\l^\a-\lb^\bb\Pi^a T_{a\bb}{}^\a,\quad Q\Pib^\a=\Nb\l^\a-\lb^\bb\Pib^a T_{a\bb}{}^\a,\cr
&Q\Pi^\ab=\N\lb^\ab-\l^\b\Pi^a T_{a\b}{}^\ab,\quad Q\Pib^\ab=\Nb\lb^\ab-\l^\b\Pib^a T_{a\b}{}^\ab .\cr
}}
Note that any superfield $\Psi$ transforms like $Q\Psi=\l^\a\N_\a\Psi+\lb^\ab\N_\ab\Psi$, again up to a Lorentz transformation with field-dependent parameter.
It is direct to verify that the transformations \QC\ are nilpotent for all the world-sheet fields, except for the $\o$ and $\ob$ variables which $Q^2$ on them give a gauge transformation \ChandiaSTA. 

Below we construct vertex operators, both unintegrated and integrated. They are on-shell objects, then we will need the equations of motion derived from the action \SC. By varying this action respect to $d, \dd$ and the pure spinor variables we obtain 
\eqn\eqsA{\eqalign{&\Pib^\a+\dd_\ab P^{\a\ab}+\lb^\ab\ob_\bb\CC_\ab{}^{\bb\a}=0,\quad ~~~~~~~\Pi^\ab-d_\a P^{\a\ab}+\l^\a\o_\b C_\a{}^{\b\ab}=0,\cr &\Nb\l^\a+\l^\b (\dd_\ab C_\b{}^{\a\ab}+\lb^\ab\ob_\bb S_{\b\ab}{}^{\a\bb} )=0, \Nb\o_\a-(\dd_\ab C_\a{}^{\b\ab}+\lb^\ab\ob_\bb S_{\a\ab}{}^{\b\bb})\o_\b=0,\cr &\N\lb^\ab+\lb^\bb(d_\a\CC_\bb{}^{\ab\a}+\l^\a\o_\b S_{\a\bb}{}^{\b\ab})=0, \N\ob_\ab-(d_\a\CC_\ab{}^{\bb\a}+\l^\a\o_\b S_{\a\ab}{}^{\b\bb})\ob_\bb=0 .}}
By varying the action \SC\ respect to $Z^M$ we obtain the equation
\eqn\eqZ{\eqalign{&-\ha E_M{}^a (\N\Pib_a+\Nb\Pi_a)+E_M{}^\a\Nb d_\a+E_M{}^\ab\N\dd_\ab-\ha\Pi^{(a}\Pib^{A)}T_{AMa}+\ha\Pi^A\Pib^BH_{BAM} \cr &+\Pib^A T_{AM}{}^\a d_\a+\Pi^A T_{AM}{}^\ab~\dd_\ab-\l^\a\o_\b\Pi^A R_{AM\a}{}^\b-\Pi^A\lb^\ab\ob_\bb R_{AM\ab}{}^\bb+d_\a\dd_\ab\N_M P^{\a\ab} \cr&+(-1)^M\l^\a\o_\b\dd_\ab\N_MC_\a{}^{\b\ab}+(-1)^Md_\a\lb^\ab\ob_\bb\N_M\CC_\ab{}^{\bb\a}+\l^\a\o_\b\lb^\ab\ob_\bb\N_MS_{\a\ab}{}^{\b\bb} =0 .\cr}}
Note that the background fields in \hol\ and \ahol\ were used. The equation for $d$ is obtained after multiplying \eqZ\ from the left by $E_\a{}^M$ and it gives 
\eqn\Nd{\eqalign{\Nb d_\a=&-d_\b\dd_\bb\N_\a P^{\b\bb}+\l^\b\o_\g(\Pib^aR_{a\a\b}{}^\g+\Pib^\bb R_{\bb\a\b}{}^\g)\cr&+\l^\b\o_\g\dd_\bb(\N_\a C_\b{}^{\g\bb}-P^{\d\bb}R_{\d\a\b}{}^\g)+d_\b\lb^\bb\ob_\gb(\N_\a \CC_\bb{}^{\gb\b}+P^{\b\db}R_{\db\a\bb}{}^\gb)\cr&-\l^\b\o_\g\lb^\bb\ob_\gb(\N_\a S_{\b\bb}{}^{\g\gb}+C_\b{}^{\g\db}R_{\db\a\bb}{}^\gb+\CC_\bb{}^{\gb\d}R_{\d\a\b}{}^\g) .\cr    }  }
Similarly by multiplying from the left with $E_M{}^\ab$ we obtain
\eqn\Ndd{\eqalign{\N \dd_\ab=&-d_\b\dd_\bb\N_\ab P^{\b\bb}+\lb^\bb\ob_\gb(\Pi^aR_{a\ab\bb}{}^\gb+\Pi^\b R_{\b\ab\bb}{}^\gb)\cr&+\l^\b\o_\g\dd_\bb(\N_\ab C_\b{}^{\g\bb}-P^{\d\bb}R_{\d\ab\b}{}^\g)+d_\b\lb^\bb\ob_\gb(\N_\ab \CC_\bb{}^{\gb\b}+P^{\b\db}R_{\db\ab\bb}{}^\gb)\cr&-\l^\b\o_\g\lb^\bb\ob_\gb(\N_\ab S_{\b\bb}{}^{\g\gb}+C_\b{}^{\g\db}R_{\db\ab\bb}{}^\gb+\CC_\bb{}^{\gb\d}R_{\d\ab\b}{}^\g) .\cr    }  }
As a check, it is direct to confirm from here, after using the equations for $\l$ and $\lb$ in \eqsA, that $\Nb(\l^\a d_\a)=\N(\lb^\ab\dd_\ab)=0$. Finally, the equations for $\Pi^A$ are obtained as follows. The identity 
\eqn\iden{\N\Pib^A-\Nb\Pi^A=-\Pi^B\Pib^C T_{CB}{}^A ,}
is necessary. The equations for $\Pi^a$ and $\Pib^a$ are obtained combining this identity with the equation \eqZ\ multiplied by $E_a{}^M$ from the left. We obtain
\eqn\Npib{\eqalign{\N\Pib_a&=\Pi^b\Pib^c T_{abc}-\Pi^\a\Pib^b T_{\a ab}+d_\a\Pib^b T_{ab}{}^\a+\Pi^b\dd_\ab T_{ab}{}^\ab-\Pi^\b\dd_\ab T_{a\b}{}^\ab \cr&+\l^\a\o_\b\Pib^bR_{ab\a}{}^\b+\Pi^b\lb^\ab\ob_\bb R_{ab\ab}{}^\bb+\Pi^\g\lb^\ab\ob_\bb R_{a\g\ab}{}^\bb+d_\a\dd_\ab\N_a P^{\a\ab}\cr&+\l^\a\o_\b\dd_\gb(\N_a C_\a{}^{\b\gb}-P^{\d\gb}R_{a\d\a}{}^\b)+d_\g\lb^\ab\ob_\bb(\N_a \CC_\ab{}^{\bb\g}+P^{\g\db}R_{a\db\ab}{}^\bb)\cr&+\l^\a\o_\b\lb^\ab\ob_\bb(\N_a S_{\a\ab}{}^{\b\bb}-C_\a{}^{\b\gb} R_{a\gb\ab}{}^\bb-\CC_\ab{}^{\bb\g} R_{a\g\a}{}^\b) ,\cr   }}
and
\eqn\Nbpi{\eqalign{\Nb\Pi_a&=\Pi^b\dd_\ab(T_{ab}{}^\ab+P^{\a\ab}T_{\a ab})+\Pi^b\lb^\ab\ob_\bb(R_{ab\ab}{}^\bb+\CC_\ab{}^{\bb\a}T_{\a ab})+d_\a\Pib^b T_{ab}{}^\a\cr&+d_\a\Pib^\bb T_{a\bb}{}^\a+\l^\a\o_\b\Pib^bR_{ab\a}{}^\b+\l^\a\o_\b\Pib^b R_{ab\a}{}^\b+\l^\a\o_\b\Pib^\gb R_{a\gb\a}{}^\b+d_\a\dd_\ab\N_a P^{\a\ab}\cr&+\l^\a\o_\b\dd_\gb(\N_a C_\a{}^{\b\gb}-P^{\d\gb}R_{a\d\a}{}^\b)+d_\g\lb^\ab\ob_\bb(\N_a \CC_\ab{}^{\bb\g}+P^{\g\db}R_{a\db\ab}{}^\bb)\cr&+\l^\a\o_\b\lb^\ab\ob_\bb(\N_a S_{\a\ab}{}^{\b\bb}-C_\a{}^{\b\gb} R_{a\gb\ab}{}^\bb-\CC_\ab{}^{\bb\g} R_{a\g\a}{}^\b) .\cr   }}
Finally, the equations for $(\Pi^\a,\Pib^\a, \Pi^\ab, \Pib^\ab)$ come from using \iden\ and the fact that $(\Pib^\a, \Pi^\ab)$ depend on other world-sheet fields according to \eqsA.

\newsec{The unintegrated vertex}

In this section we study the fluctuations around the background of the previous section. We start with the unintegrated vertex operator. As in flat space it takes the simple form
\eqn\UC{U=\l^\a\lb^\bb \AB_{\a\bb}(Z) ,}
but now the background is not flat anymore. Imposing that this vertex is in the cohomology of the BRST charge determines the equations
\eqn\eqa{\N_{(\a}\AB_{\b)\gb}=(\g^a)_{\a\b}\AB_{a\gb},\quad \N_{(\ab}\AB_{\g\bb)}=(\g^a)_{\ab\bb}\AB_{\g a} ,}
which look similar to \Ams. We proceed as in flat space. That is, we obtain constraining equations for superfields defined after taking higher $\N_\a$ and $\N_\ab$ derivatives on $\AB_{\a\ab}$. 

As the first step, we get equations similar to \DL\ and \DR\ of flat space. Consider the first equation in \eqa. It can be shown that 
\eqn\eqb{\g^a_{(\a\b}\left(\N_{\g)}\AB_{a\ab}-2\O_{\g)}\AB_{a\ab} - \N_a\AB_{\g)\ab}\right)=0 ,}
which implies the existence of a superfield $\WB^\b{}_\ab$ satisfying
\eqn\eqc{\N_\a\AB_{a\ab}-2\O_\a\AB_{a\ab}-\N_a\AB_{\a\ab}=(\g_a)_{\a\b}\WB^\b{}_\ab .}
This is equivalent to the first equation in \DL\ of flat space-time background. The equation \eqb\ is verified by plugging $A_{a\ab}$ from the first in \eqa\ and then commuting the covariant derivatives as
\eqn\comm{[\N_A,\N_B]\MB_{C\cdots}{}^{D\cdots}=-T_{AB}{}^E\N_E\MB_{C\cdots}{}^{D\cdots}+\MB_{C\cdots}{}^{E\cdots}R_{ABE}{}^D-R_{ABC}{}^E\MB_{E\cdots}{}^{D\cdots}+\cdots,}
where $\MB$ is a tensor. The torsion and curvature components given in \hol\ and \ahol\ are also necessary.

To get the analogous to the second equation in \DL\ we note that there is a superfield $\Psi_\a{}^\b{}_\gb$, containing $\N_\a\WB^\b{}_\gb$, that satisfies an equation involving $\Psi_\a{}^\b{}_\gb+{1\over{10}}\g^a_{\a\r}\g_a^{\b\d}\Psi_\d{}^\r{}_\gb$. It turns out that
\eqn\eqpsi{\eqalign{\Psi_\a{}^\b{}_\gb&=\N_\a\WB^\b{}_\gb-4\O_\a\WB^\b{}_\gb+P^{\b\rb}\N_\rb \AB_{\a\gb}-\CC_\gb{}^{\rb\b}\AB_{\a\rb}\cr&+2(\g^a)^{\b\r}(\N_\a\O_\r \AB_{a\gb}-\O_\r\N_a \AB_{\a\gb} ) ,}}
satisfies the equation
\eqn\eqpsip{\eqalign{\Psi_\a{}^\b{}_\gb+{1\over{10}}\g^a_{\a\r}\g_a^{\b\d}\Psi_\d{}^\r{}_\gb&={1\over{10}}(\g^{ab})_\a{}^\b\left(\FB_{ab\gb}+T_{ab}{}^c\AB_{c\gb}+T_{ab}{}^\r \AB_{\r\gb}+3\tau_{ab}{}^c\AB_{c\gb}\right)\cr&-{1\over{10}}(\g^{abcd})_\a{}^\b\tau_{abc}\AB_{d\gb},\cr}}
where $\FB_{ab\gb}=\N_{[a}\AB_{b]\gb}$ and $\tau_{abc}=\g_{abc}^{\a\b}\O_\a\O_\b$. Note this type of equation was obtained in \ChandiaNYH\ for the heterotic string in curved background. The solution of \eqpsip\ is
\eqn\psiis{\Psi_\a{}^\b{}_\gb={1\over4}(\g^{ab})_\a{}^\b \left( \FB_{ab\gb}+T_{ab}{}^c\AB_{c\gb}+T_{ab}{}^\r \AB_{\r\gb}+3\tau_{ab}{}^c\AB_{c\gb} \right) - {1\over12} (\g^{abcd})_\a{}^\b \tau_{abc}\AB_{d\gb} .}
Similarly, from the second equation in \eqa\ we obtain
\eqn\Wb{\N_\ab\AB_{\g a}-2\Ob_\ab\AB_{\g a}-\N_a\AB_{\g\ab}=(\g_a)_{\ab\bb}\WB_\g{}^\bb,}
and the combinations of fields
\eqn\psib{\eqalign{\Psib_{\g\ab}{}^\bb&=\N_\ab\WB_\g{}^\bb-4\Ob_\ab\WB_\g{}^\bb-P^{\r\bb}\N_\r \AB_{\g\ab}-C_\g{}^{\r\bb}\AB_{\r\ab}\cr&+2(\g^a)^{\bb\rb}(\N_\ab\Ob_\rb \AB_{\g a}-\Ob_\rb\N_a \AB_{\g\ab} ) ,}}
turns out to be equal to 
\eqn\psibis{\Psib_{\g\ab}{}^\bb={1\over4}(\g^{ab})_\ab{}^\bb \left( \FB_{\g ab}+T_{ab}{}^c\AB_{\g c}+T_{ab}{}^\rb \AB_{\g\rb}+3\taub_{ab}{}^c\AB_{\g c} \right) - {1\over{12}} (\g^{abcd})_\ab{}^\bb \taub_{abc}\AB_{\g d} ,}
where $\FB_{\g ab}=\N_{[a}\AB_{\g b]}$ and $\taub_{abc}=\g_{abc}^{\ab\bb}\Ob_\ab\Ob_\bb$.

In the next section we construct the integrated vertex operator from the chain of superfields $(\AB_{\a\ab}, \AB_{a\ab},Ê\AB_{\a a}, \WB^\a{}_\ab, \WB_\a{}^\ab)$ just like we did in section $2$ for the flat space-time background.

\newsec{The integrated vertex}

The integrated vertex operator is given by the same descent procedure of flat space-time background. There, the existence of a composite $b$ ghost was crucial to state that any operator with conformal weight different from zero is trivial in the cohomology of the BRST charge. In particular, $\p U=QW$ and $\pb U=Q\WW$. Similarly, $\p\WW-\pb W$ is also trivial, therefore it is equal to $QV$ with $V$ being the integrated vertex operator \ivoz. For the generic type II supergravity background of section $3$, we need that the $b$ ghosts of left- and right-moving sectors. Such composite $b$ ghost exists for the heterotic string in a generic background field \ChandiaIMA\ \BerkovitsAMA. Let us assume that the composite $b$ ghosts exist for the type II superstring in a generic background. 

A $W$ satisfying $QW=\p U$  is given by 
\eqn\Wis{\eqalign{W&=\lb^\bb [\Pi^\a \AB_{\a\bb}+\Pi^a\AB_{a\bb}+d_\a(\WB^\a{}_\bb+2(\g^a)^{\a\g}\O_\g\AB_{a\bb})\cr &-J \O_\r \WB^\r{}_\bb+\ha N^{ab}(\FB_{ab\bb}+T_{ab}{}^c \AB_{c\bb}+T_{ab}{}^\r\AB_{\r\bb}+4\tau_{ab}{}^c\AB_{c\bb}+2(\O\g_{ab}\WB_\bb) ) ] ,}}
where $J=\l^\a\o_\a$ and $N^{ab}=\ha(\l\g^{ab}\o)$.
And a $\WW$ satisfying $Q\WW=\pb U$ is given by
\eqn\Wbis{\eqalign{\WW&=\l^\a [\Pib^\bb \AB_{\a\bb}+\Pib^a\AB_{\a a}+\dd_\bb(\WB_\a{}^\bb+2(\g^a)^{\bb\gb}\Ob_\gb\AB_{\a a})\cr &-\Jb \Ob_\rb \WB_\a{}^\rb+\ha \NN^{ab}(\FB_{\a ab}+T_{ab}{}^c \AB_{\a c}+T_{ab}{}^\rb\AB_{\a\rb}+4\taub_{ab}{}^c\AB_{\a c}+2(\Ob\g_{ab}\WB_\a) ) ] ,}}
where $\Jb=\lb^\ab\ob_\ab$ and $\NN^{ab}=\ha(\lb\g^{ab}\ob)$.

The integrated vertex operator satisfying $QV=\p\WW-\pb W$ has the form
\eqn\Vis{\eqalign{V&=\Pi^\a\Pib^\ab \AB_{\a\ab}+\Pi^\a\Pib^a \AB_{\a a}-\Pi^a\Pib^\ab \AB_{a\ab}+\Pi^\a\dd_\ab\Phib_\a{}^\ab+ d_\a\Pib^\ab\Phib_\ab{}^\a+\Pi^\a\lb^\ab\ob_\bb\Phib_{\a\ab}{}^\bb\cr&+\l^\a\o_\b\Pib^\ab\Phib_{\ab\a}{}^\b+\Pi^a\Pib^b\AB_{ab}+d_\a\dd_\ab \PB^{\a\ab}+\Pi^a\dd_\ab\EB_a{}^\ab+d_\a\Pib^a \EBB_a{}^\a+\Pi^a\lb^\ab\ob_\bb\OB_{a\ab}{}^\bb\cr&+\l^\a\o_\b\Pib^a\OBB_{a\a}{}^\b+d_\a\lb^\ab\ob_\bb\CBB_\ab{}^{\bb\a}+\l^\a\o_\b\dd_\ab\CB_\a{}^{\b\ab}+\l^\a\o_\b\lb^\ab\ob_\bb\SB_{\a\ab}{}^{\b\bb} ,\cr}}
where
\eqn\phis{\eqalign{&\Phib_\a{}^\ab=\WB_\a{}^\ab+2(\g^a\Oh)^\ab \AB_{\a a},\quad \Phib_\ab{}^\a=\WB^\a{}_\ab+2(\g^a\O)^\a \AB_{a\ab},\cr &\Phib_{\a\ab}{}^\bb=-\d_\ab^\bb\Oh_\gb\WB_\a{}^\gb+{1\over4}(\g^{ab})_\ab{}^\bb\left( \FB_{\a ab}+T_{ab}{}^c\AB_{\a c}+T_{ab}{}^\bb\AB_{\a\bb}+4\taub_{ab}{}^c\AB_{\a c}+2(\Oh\g_{ab}\WB_\a)  \right)   ,\cr&\Phib_{\ab\a}{}^\b=\d_\a^\b\O_\g\WB^\g{}_\ab-{1\over4}(\g^{ab})_\a{}^\b\left(\FB_{ab\ab}+T_{ab}{}^c\AB_{c\ab}+T_{ab}{}^\b\AB_{\b\ab}+4\tau_{ab}{}^c\AB_{c\ab}+2(\O\g_{ab}\WB_\ab) \right)   .\cr }}
$\AB_{ab}$ is defined by 
\eqn\ABab{\eqalign{&\N_{(\a}\AB_{\b)a}+T_{(\a a}{}^b\AB_{\b)b}-T_{a(\a}{}^\bb\AB_{\b)\bb}=\g^b_{\a\b}\AB_{ba},\cr&\N_{(\ab}\AB_{a\bb)}-T_{a(\ab}{}^\b\AB_{\b\bb)}=-\g^b_{\ab\bb}\AB_{ab} .}}
Note that the second equation here does not contain the torsion component $T_{\ab a}{}^b$ because it vanishes. The fluctuations $\EB_a{}^\ab$ and $\EBB_a{}^\a$ are given by
\eqn\EBs{\eqalign{&\N_{(\a}\left(\WB_{\b)}{}^\ab+2(\g^a\Oh)^\ab\AB_{\b)a}\right)+T_{a(\a}{}^\ab\AB_{\b)}{}^a=-\g^a_{\a\b}\EB_a{}^\ab,\cr&\N_{(\ab}\left(\WB^\a{}_{\bb)}+2(\g^a\O)^\a\AB_{b\bb)}\right)+T_{a(\ab}{}^\a\AB^a{}_{\bb)}=\g^a_{\ab\bb}\EBB_a{}^\a .\cr }}
$\OB_{a\ab}{}^\bb$ and $\OBB_{a\a}{}^\b$ are given by
\eqn\OBs{\eqalign{&\N_{(\a} \Phib_{\b)\ab}{}^\bb -R_{\gb(\a\ab}{}^\bb\Phib_{\b)}{}^\gb-R_{a(\a\ab}{}^\bb \AB_{\b)}{}^a=\g^a_{\a\b}\OB_{a\ab}{}^\bb, \cr & \N_{(\ab}\Phib_{\bb)\a}{}^\b+R_{\g(\ab\a}{}^\b\Phib_{\bb)}{}^\g-R_{a(\ab\a}{}^\b\AB^a{}_{\bb)}=\g^a_{\ab\bb}\OBB_{a\a}{}^\b .\cr}}
$\PB$ is given by
\eqn\PBis{\eqalign{&\N_\a\EB_a{}^\ab+\N_a\Phib_\a{}^\ab- T_{\a a}{}^b\EB_b{}^\ab -T_{ab}{}^\ab\AB_\a{}^b-T_{\a b}{}^\ab\AB_a{}^b-(P\g_aP)^{\ab\bb}\AB_{\a\bb}=-(\g_a)_{\a\b}\PB^{\b\ab}  \cr & \N_\ab\EBB_a{}^\a-\N_a\Phib_\ab{}^\a+T_{ab}{}^\a\AB^b{}_\ab-T_{\ab b}{}^\a\AB^b{}_a+(P\g_aP)^{\a\b}\AB_{\b\ab}=-(\g_a)_{\ab\bb}\PB^{\a\bb}  .\cr}}
$\CBB_\ab{}^{\bb\a}$ and $\CB_\a{}^{\b\ab}$ are given by
\eqn\CBis{\eqalign{&\N_\a\OB_{a\ab}{}^\bb-\N_a\Phib_{\a\ab}{}^\bb+T_{\a a}{}^b\OB_{\b\a}{}^\b-R_{\a b\ab}{}^\bb\AB_a{}^b+R_{\a\gb\ab}{}^\bb\EB_a{}^\gb\cr&-R_{a\gb\ab}{}^\bb\Phib_\a{}^\gb+R_{ab\ab}{}^\bb\AB_\a{}^b+\CC_\ab{}^{\bb\g}T_{a\g}{}^\gb\AB_{\a\gb}=(\g_a)_{\a\b}\CBB_\ab{}^{\bb\b}  ,\cr &\N_\ab\OBB_{a\a}{}^\b-\N_a\Phib_{\ab\a}{}^\b-R_{\ab b\a}{}^\b\AB^b{}_a+R_{\ab\g\a}{}^\b\EBB_a{}^\g\cr&+R_{a\g\a}{}^\b\Phib_\ab{}^\g-R_{ab\a}{}^\b\AB^b{}_\ab-\CC_\a{}^{\b\gb}T_{a\gb}{}^\g\AB_{\g\ab}=-(\g_a)_{\ab\bb}\CB_\a{}^{\b\bb}  .\cr }}
And finally, $\SB$ is given by
\eqn\SBis{\eqalign{&\SB_{\a\ab}{}^{\b\bb}=\N_\a\CBB_\ab{}^{\bb\b}+R_{\a\gb\ab}{}^\bb\PB^{\b\gb}-R_{\a a\ab}{}^\bb\EBB^{a\b}+\left(\N_a\CC_\ab{}^{\bb\b}+P^{\g\db}R_{a\db\ab}{}^\bb\right)\AB_\a{}^a \cr &+\left(\N_\gb\CC_\ab{}^{\bb\b}+P^{\b\db}R_{\db\gb\ab}{}^\bb\right)\Phib_\a{}^\gb +P^{\b\gb}\N_\gb\Phib_{\a\ab}{}^\bb+\CC_\ab{}^{\gb\b}\Phib_{\a\gb}{}^\bb-\CC_\gb{}^{\bb\b}\Phib_{\a\ab}{}^\gb  \cr}} 
or
\eqn\SBisp{\eqalign{&\SB_{\a\ab}{}^{\b\bb}=\N_\ab\CB_\a{}^{\b\bb}-R_{\ab\g\a}{}^\b\PB^{\g\bb}-R_{\ab a\a}{}^\b\EB^{a\bb}-\left(\N_a C_\a{}^{\b\bb}-P^{\d\gb}R_{a\d\a}{}^\b\right)\AB^a{}_\ab \cr &-\left(\N_\g C_\a{}^{\b\bb}-P^{\d\bb}R_{\d\g\a}{}^\b\right)\Phib_\ab{}^\g -P^{\g\bb}\N_\g\Phib_{\ab\a}{}^\b-C_\a{}^{\g\bb}\Phib_{\ab\g}{}^\b+C_\g{}^{\b\bb}\Phib_{\ab\a}{}^\g .\cr}} 

The physical interpretation of the fluctuations is similar to the flat space-time case. The operator $\AB_{ab}$ of \ABab\ describes fluctuations of the metric, the Kalb-Ramond and the dilaton superfields. The operators $\EB_a{}^\ab$ and $\EBB_a{}^\a$ in \EBs\ describe fluctuations of the vielbein superfields. The operators $\OB_{a\ab}{}^\bb$ and $\OBB_{a\a}{}^\b$ describe fluctuations of the Lorentz connection superfields. The operator $\PB^{\a\bb}$ of \PBis\ describes the fluctuation of the Ramond-Ramond field strength. 

It is interesting to note that if the fluctuation \Vis\ is added to the world-sheet action \SC\ give a new supergravity background (linear in the fluctuations). Note that this was also the case for the heterotic string in a generic background \ChandiaNYH. The action becomes
\eqn\Smod{\eqalign{S&=\int d^2z\ha\p Z^M \pb Z^N (G'_{NM}+B'_{NM})+d_\a \pb Z^M E'_M{}^\a+\dd_\ab\p Z^M E'_M{}^\ab \cr &+\l^\a\o_\b\p Z^M\O'_{M\a}{}^\b+\lb^\ab\ob_\bb\p Z^M\O'_{M\ab}{}^\bb+d_\a\dd_\ab P'^{\a\ab}+\l^\a\o_\b\dd_\ab C'_\a{}^{\b\ab} \cr&+\lb^\ab\ob_\b d_\a\CC'_\ab{}^{\bb\a}+\l^\a\o_\b\lb^\ab\ob_\bb S'_{\a\ab}{}^{\b\bb}+\o_\a\p\l^\a+\ob_\ab\p\lb^\ab ,\cr }}
which is equal to the action \SC\ if we remove the primes. Note that the supermetric $G_{NM}$ is equal to $E_N{}^a E_{Ma}$. The explicit expressions for the primed superfields in terms of unprimed superfields is as follows 
\eqn\prupr{\eqalign{&G'_{NM}=G_{NM}+ [ (-1)^{M+1}E_N{}^\ab E_M{}^\a\AB_{\a\ab}+E_N{}^a E_M{}^\a\AB_{\a a}+(-1)^M E_N{}^\ab E_M{}^a\AB_{a\ab}\cr &~~~~~~~~~~~~~~~~~~~~~+E_N{}^b E_M{}^a \AB_{ab}+(M \leftrightarrow N)  ] ,\cr &B'_{NM}=B_{NM} + [ (-1)^{M+1}E_N{}^\ab E_M{}^\a\AB_{\a\ab}+E_N{}^a E_M{}^\a\AB_{\a a}+(-1)^M E_N{}^\ab E_M{}^a\AB_{a\ab}\cr &~~~~~~~~~~~~~~~~~~~~~+E_N{}^b E_M{}^a \AB_{ab}-(M \leftrightarrow N)  ]  ,\cr&E'_M{}^\a=E_M{}^\a+E_M{}^\ab\Phib_\ab{}^\a+E_M{}^a\EBB_a{}^\a ,\cr&E'_M{}^\ab=E_M{}^\ab-E_M{}^\a\Phib_\a{}^\ab+E_M{}^\ab\EB_a{}^\ab ,\cr &\O'_{M\a}{}^\b=\O_{M\a}{}^\b+E_M{}^\ab\Phib_{\ab\a}{}^\b+E_M{}^a\OBB_{a\a}{}^\b ,\cr&\O'_{M\ab}{}^\bb=\O_{M\ab}{}^\bb+E_M{}^\a\Phib_{\a\ab}{}^\bb+E_M{}^a\OB_{a\ab}{}^\bb ,\cr& P'^{\a\ab}=P^{\a\ab}+\PB^{\a\ab} ,\cr&C'_\a{}^{\b\ab}=C_\a{}^{\b\ab}+\CB_\a{}^{\b\ab} ,\cr&\CC'_{\ab}{}^{\bb\a}=\CC_\ab{}^{\bb\a}+\CBB_\ab{}^{\bb\a} ,\cr&S'_{\a\ab}{}^{\b\bb}=S_{\a\ab}{}^{\b\bb}+\SB_{\a\ab}{}^{\b\bb}  .\cr }} 

\newsec{Gauge symmetry}

Up to now we have studied the consequences of having the unintegrated vertex operator \UC\ which is annihilated by the pure spinor BRST charge. Note that $U$ has ghost-number $(1,1)$ and vanishing conformal dimension. By the descent procedure used above, there exist world-sheet fields $W$ (of ghost number $(0,1)$ and conformal dimension $(1,0)$) and $\WW$ (of ghost number $(1,0)$ and conformal dimension $(0,1)$) satisfying $QW=\p U$ and $Q\WW=\pb U$, respectively. And finally, the integrated vertex operator $V$, of vanishing ghost number and conformal dimension $(1,1)$, satisfies $QV=\p\WW-\pb W$. The unintegrated vertex is not only annihilated by $Q$, it belongs to the cohomology of $Q$, that is $U$ is also defined up to $Q\L$ for some $\L$. On generic grounds, $W$ is defined up to $\p\L + Q\S$ for some $\S$ and $\WW$ is defined up to $\pb\L+Q{\overline\S}$ for some ${\overline\S}$. Finally, the integrated vertex operator is defined up to $\p{\overline\S}-\pb\S$, that is a total derivative. Note that an BRST-exact term is not allowed in the transformation of the integrated vertex operator because $Q$ should act on a field with negative ghost number. This is not possible in the pure spinor formalism.This construction was explicitly done in \ChandiaAFC\ for the type II superstring in a flat background. We now consider our case.

The gauge invariance $\d U=Q\L$ implies a gauge transformation for the superfield $\AB_{\a\bb}$. Giving $\L=\l^\a\LB_\a+\lb^\ab\LBB_\ab$, the gauge transformation of $\AB_{\a\bb}$ is
\eqn\dAab{\d\AB_{\a\bb}=\N_\a\LBB_\bb+\N_\bb\LB_\a ,}
where the gauge parameters have to satisfy
\eqn\eqL{\N_{(\a}\LB_{\b)}=\g^a_{\a\b}\LB_a,\quad \N_{(\ab}\LBB_{\bb)}=\g^a_{\ab\bb}\LBB_a,}
in order to preserve the form \UC\ of the unintegrated vertex operator. Note that the equations in \eqL\ have a structure similar to the equations \eqa, then we should have equations like \eqb, \eqc, \psiis\ and \psibis. In fact, performing the kind of calculations of the section 4 we obtain
\eqn\eqLL{\eqalign{&\N_\a\LB_a-2\O_\a\LB_a-\N_a\LB_\a=(\g_a)_{\a\b}\LB^\b, \cr&\N_\a\LB^\b-4\O_\a\LB^\b+P^{\b\gb}\N_\gb\LB_\a+2(\g^a)^{\b\g}(\N_\a\O_\g \LB_a-\O_\g\N_a\LB_\a)\cr&={1\over4}(\g^{ab})_\a{}^\b\left( \N_{[a}\LB_{b]}+T_{ab}{}^c\LB_c+T_{ab}{}^\g\LB_\g+3\tau_{ab}{}^c\LB_c \right)-{1\over{12}}(\g^{abcd})_\a{}^\b\tau_{abc}\LB_d ,\cr
&\N_\ab\LBB_a-2\Ob_\ab\LBB_a-\N_a\LBB_\ab=(\g_a)_{\ab\bb}\LBB^\bb, \cr&\N_\ab\LBB^\bb-4\Ob_\ab\LBB^\bb-P^{\g\bb}\N_\g\LBB_\ab+2(\g^a)^{\bb\gb}(\N_\ab\Ob_\gb \LBB_a-\Ob_\gb\N_a\LBB_\ab)\cr&={1\over4}(\g^{ab})_\ab{}^\bb\left( \N_{[a}\LBB_{b]}+T_{ab}{}^c\LBB_c+T_{ab}{}^\gb\LBB_\gb+3\taub_{ab}{}^c\LBB_c \right)-{1\over{12}}(\g^{abcd})_\ab{}^\bb\taub_{abc}\LBB_d ,\cr}}
where $\tau_{abc}=\g_{abc}^{\a\b}\O_\a\O_\b$  and $\taub_{abc}=\g_{abc}^{\ab\bb}\Ob_\ab\Ob_\bb$.

We now find the gauge transformations for the remaining superfields in $W$. Performing a gauge transformation in the first equation of \eqa\ and in \eqc\ we obtain
\eqn\dmore{\eqalign{&\d\AB_{a\ab}=\N_a\LBB_\ab-\N_\ab\LB_a-(\g_aP)_\ab{}^\a\LB_\a,\cr &\d\WB^\a{}_\ab=\N_\ab(\LB^\a-P^{\a\bb}\LBB_\bb)-(P\g^a)^\a{}_\ab(\LB_a-\LBB_a)\cr&+2\N_\ab\left((\g^a\O)^\a\right)\LB_a+2(\g^a\O)^\a\left(\N_a\LBB_\ab+(\g_aP)_\ab{}^\b\LB_\b\right) . \cr}}
Using these transformations in \Wis\ one obtains that its gauge transformation becomes
\eqn\dWis{\eqalign{&\d W=\p(\l^\a\LB_\a+\lb^\bb\LBB_\ab)-Q (\Pi^\a\LB_\a+\Pi^a\LB_a+d_\a(\LB^\a+2(\g^a\O)^\a\LB_a) \cr&-J\O_\a\LB^\a+\ha N^{ab}(\N_{[a}\LB_{b]}+T_{ab}{}^c\LB_c+T_{ab}{}^\a\LB_\a+4\tau_{ab}{}^c\LB_c+2(\O\g_{ab})^\a\LB_\a ) ,\cr}}
as expected from the above discussion. Similarly, for the superfields in $\WW$ we obtain
\eqn\dmore{\eqalign{&\d\AB_{a\ab}=\N_a\LBB_\ab-\N_\ab\LB_a-(\g_aP)_\ab{}^\a\LB_\a,\cr &\d\WB^\a{}_\ab=\N_\ab(\LB^\a-P^{\a\bb}\LBB_\bb)-(P\g^a)^\a{}_\ab(\LB_a-\LBB_a)\cr&+2\N_\ab\left((\g^a\O)^\a\right)\LB_a+2(\g^a\O)^\a\left(\N_a\LBB_\ab+(\g_aP)_\ab{}^\b\LB_\b\right) . \cr}}
Using these transformations in \Wbis\ one obtains
\eqn\dWis{\eqalign{&\d\WW=\pb(\l^\a\LB_\a+\lb^\bb\LBB_\ab)-Q (\Pib^\ab\LBB_\ab+\Pib^a\LBB_a+\dd_\ab(\LBB^\ab+2(\g^a\Oh)^\ab\LBB_a) \cr&-\Jb\Oh_\ab\LBB^\ab+\ha \NN^{ab}(\N_{[a}\LBB_{b]}+T_{ab}{}^c\LBB_c+T_{ab}{}^\ab\LBB_\ab+4\taub_{ab}{}^c\LBB_c+2(\Oh\g_{ab})^\ab\LBB_\a ) ,\cr}}
as expected from above.

Finally, the integrated vertex operator transforms to a total derivative. In fact,
\eqn\dVis{\eqalign{&\d V=\pb(\Pi^\a\LB_\a+\Pi^a\LB_a+d_\a(\LB^\a+2(\g^a\O)^\a\LB_a) \cr&-J\O_\a\LB^\a+\ha N^{ab}(\N_{[a}\LB_{b]}+T_{ab}{}^c\LB_c+T_{ab}{}^\a\LB_\a+4\tau_{ab}{}^c\LB_c+2(\O\g_{ab})^\a\LB_\a ) \cr
&-\p(\Pib^\ab\LBB_\ab+\Pib^a\LBB_a+\dd_\ab(\LBB^\ab+2(\g^a\Oh)^\ab\LBB_a) \cr&-\Jb\Oh_\ab\LBB^\ab+\ha \NN^{ab}(\N_{[a}\LBB_{b]}+T_{ab}{}^c\LBB_c+T_{ab}{}^\ab\LBB_\ab+4\taub_{ab}{}^c\LBB_c+2(\Oh\g_{ab})^\ab\LBB_\a ) .\cr}}
Again, this is what we discussed at the beginning of the current section.

\newsec{Final comments}

In this paper we have generalize the construction of the integrated vertex operator for the type II superstring in the pure spinor string in a flat background to the on-shell supergravity background. This is a generalization of the analysis done in \ChandiaKJA\ for the type IIB superstring in a $AdS_5\times S^5$ background and in  \ChandiaEKC\ for the plane wave limit of $AdS_5\times S^5$ background. We used descent procedure used to construct the integrated vertex operator. Because this procedure is based on the existence of reparametrization $b$ ghosts, there should be possible to construct these ghosts in a generic supergravity background generalizing the result of \ChandiaIMA\ and \BerkovitsAMA. It would be interesting to study the correlations among vertex operators with the hope of the determination of scattering amplitudes, we leave this problem for the  future.

\bigskip
\bigskip 
\noindent

\listrefs
 
\end